\documentstyle[epsfig]{article}

\def\spose#1{\hbox to 0pt{#1\hss}}
\def\lta{\mathrel{\spose{\lower 3pt\hbox{$\mathchar"218$}}
     \raise 2.0pt\hbox{$\mathchar"13C$}}}
\def\gta{\mathrel{\spose{\lower 3pt\hbox{$\mathchar"218$}}
     \raise 2.0pt\hbox{$\mathchar"13E$}}}

\begin{document}

\title{Radiation from Dwarf Nova Discs}
\author{Irit Idan$^1$, Jean-Pierre Lasota$^2$, Jean-Marie Hameury$^3$\\
and\\
Giora Shaviv$^4$\\
$^1${\it Rafael, Haifa, 31021 Israel}\\
$^2${\it DARC, Observatoire de Paris, 92190 Meudon, France}\\
$^3${\it Observatoire de Strasbourg, 67000 Strasbourg, France}\\
$^4${\it Department of Physics, Technion, Haifa, 32000 Israel}\\
}
\date{}
\maketitle

\begin{abstract}
We use the Shaviv \& Wehrse (1991) code to model the vertical structure
and the emission properties of quiescent dwarf nova discs. We find that
in the case of HT Cas the quiescent disc must be optically thin, in 
contradiction with the requirements of the standard disc instability 
model. We find a viscosity parameter $\alpha \gta 1$.
Although this is much less than values ($\sim 10^2$) obtained in isothermal
slab models it is not consistent with the accretion disc model
assumptions.
\end{abstract}

\noindent
\leftline{PACS: 95.10.Gz; 97.30Jx; 97.30.Qt; 97.80.Gm}
\leftline{Keywords: Accretion discs; Dwarf novae; Radiative transfer}

\section{Introduction}

Dwarf novae are cataclysmic binary systems (Warner 1995) which go into
outbursts at more or less regular intervals. In cataclysmic binaries, a
Roche--lobe filling low mass (secondary) star loses matter that is
accreted by a white dwarf (the primary).  If the white dwarf is not too
strongly magnetized ($B \lta 10^5$ G) the accreting matter forms a disc
that extends down to the white dwarf surface. Despite the variety of
the observed properties of dwarf nova outbursts it has been firmly
established that they are due to the brightening of the accretion disc
in this systems. This does not mean that the physical process leading
to outbursts must be a disc instability. For some time (see Cannizzo
1993, 1997, for the history of the subject) an instability in the  mass
transfer from the secondary was put forward as an alternative
possibility. This model was discarded by the majority of those working
in the field when a simple physical reason for  disc instability
(hydrogen recombination) was found in the early 80's. In the standard
version of the disc instability model (DIM) it is assumed that the mass
transfer is constant in time, but there is growing evidence that
variations in the mass transfer rate play an important role, in, at
least, some types of dwarf nova outbursts (Smak 1996; Lasota 1996;
Cannizzo \& King 1998).

The DIM reproduces rather well the basic properties of `typical' dwarf
nova outbursts (see Hameury et al. 1998 for the most recent version of
the model). When it comes to the description of the variety of outburst
properties, however, the range of recurrence times and the emission
properties of dwarf nova cycles, the DIM is less successful. Very long
recurrence times in WZ~Sge--type systems, are obtained only by assuming
either very low values of the viscosity parameter (Smak 1998), or inner
disc radii larger than the white dwarf radius and increased  mass
transfer during the outburst (Lasota et al. 1995; Hameury et al. 1997).
Clearly, these problems are due, at least in part, to our poor
knowledge of the viscous processes operating in accretion discs.

Additional difficulties appear when one tries to obtain accretion disc
spectra. In the simplest case, it is assumed that each ring of an
optically thick disc emits like a black body. Another approach consists
in using {\sl stellar} atmosphere models corresponding to the disc
effective temperature and gravity at optical depth one. As shown by
Wade (1988) and Shaviv \& Wehrse (1991) such models provide an
inadequate description of accretion disc spectra. In the case of
optically thin discs, various approximate schemes have been used to
describe continuum and line radiation (Williams 1980; Tylenda 1981;
Marsh 1987; Lin et al. 1988; Mineshige \& Wood 1990, Dumont et al.
1991; Williams 1991; Wood et al. 1992) but these models are not
self-consistent; in particular they often require very high ($>> 1$)
values of the viscosity parameter (see below).

Quiescent dwarf nova discs present a rather surprising puzzle, which
results, probably, from two weaknesses of accretion disc models: the
ignorance of the nature of viscosity which drives accretion
(see e.g. Gammie \& Menou 1997) and the inadequate description of the
disc emission. These two features are not independent, since the
disc temperature stratifications depend on the vertical distribution
of the viscous heating.

The disc instability model in its `standard' form predicts an optically
thick quiescent accretion disc. In several cases, however, the observed
continuum radiation of such discs suggests rather an optically thin
disc emission because of the presence of the Balmer jump in emission
and, at longer wavelengths, emission dominated by the Paschen continuum
(see e.g. Wood et al. 1989; Wood 1990; Wood et al. 1992). The simple
(isothermal slab) models of such discs, however, require enormous
values ($ \sim 10^2$) of the viscosity parameter which, according to
the model, should be smaller than unity and whose value in quiescence,
according to the DIM, should be $\sim 10^{-2}$.  Interestingly,
modeling Balmer emission lines in such discs also requires very large
values of the viscosity parameter, but in this case emission can be
attributed to an optically thin chromosphere (Mineshige \& Wood 1990).
As pointed out by Wood et al. (1992; see also Vrielmann 1997) the very
high values of the viscosity parameter most probably result from the
simplistic treatment of radiation from an optically thin discs.  There
exists, however, a {\sl self--consistent} description of the disc
vertical structure developed by Shaviv \& Wehrse (1991; hereafter SW)
and successfully applied to nova--like cataclysmic variables (Idan \&
Shaviv 1996).

In this article we apply the SW code to, non-equilibrium, dwarf nova
discs and compare the calculated emission with the general properties
of observed quiescent accretion discs in such system, in particular,
with the multicolour observations of HT Cas in quiescence (Horne et
al. 1991; Wood et al 1992).

\section{The case of HT Cas}

HT Cas is an eclipsing dwarf nova with an orbital period of 106 min.
This well--studied system is {\sl not} a typical dwarf nova: its outbursts
are rare (the mean cycle length is 400$^d\pm50^d$ according to Wenzel
1987) and although it is an SU UMa--type dwarf nova, only one
superoutburst of this system has been observed until now (Zhang et al.
1986).  In addition HT Cas also shows {\sl low
states}, i.e. during quiescence it goes to a lower luminosity state. It
is an eclipsing X--ray  source. King (1997) speculates that it could be
a cataclysmic variable that went through the minimum period and Lasota
et al. (1995) suggest that its long recurrence time could be due to the
low mass transfer rate and a truncated inner accretion disc.

Although HT Cas is an eclipsing system, almost no bright spot modulation is
observed in its light curve. This could mean (Smak, private communication)
that most of the accretion stream overflows the disc. If this is
the case, the validity of models assuming that the observed radiation
is emitted by an axially symmetric `standard' disc is questionable.
Wood (1990) used a simple models of stream penetration and stripping
to show that this does not lead to flat temperature profiles in
stationary accretion discs. Modeling multicolour properties of
quiescent dwarf-nova disc in such a configuration does not seem, however,
to be a tractable problem.

Patterson (1981) called HT Cas `the Rosetta Stone of dwarf novae' but
this system might be too atypical to provide the information needed to
solve the mysteries of dwarf novae. For the moment, however, HT Cas is
the system which provides most observations about quiescent dwarf nova
discs.

Accretion disc eclipse light curves are analyzed with the maximum
entropy method (MEM) developed by Keith Horne and collaborators (Horne
1993 and references therein). The intensity maps of the disc
produced by this technique are used to derive the surface brightness as
a function of radius.  When the distance to the system is known, the
brightness is transformed into flux density and the result is usually
represented as a {\sl brightness temperature} distribution.

The MEM technique was applied to quiescent disc eclipse light curves of
Z Cha (Wood el al. 1986), OY Car (Rutten el al. 1992; Wood et al. 1989)
and HT Cas (Wood et al. 1992). Only in the case of HT Cas the technique
was applied to light curves in four  UBVR colours. In all cases, as
predicted by the DIM (see e.g. Smak 1984, Fig 4). the radial
distribution of temperature is flat, indicating a non--steady disc.
The DIM predicts a flat profile of the {\sl effective temperature} (see
below) and the brightness temperature is close to the effective one
only if the disc radiates approximatly like a blackbody - which is not
the case even for optically thick accretion discs (see e.g. SW) In
Z Cha, OY Car and HT Cas, however, the disc colours suggest that it is
optically thin so care should be taken when interpreting the MEM results.

The case of HT Cas seem to be more complicated. Zhang et al. (1996)
found that the disc (observed between October 1982 and October 1984) is
optically thick but the October-November 1983 UBVR observations
(with one observation the same night as Zhang et al.)  by
Horne et al. (1991) show that disc emission is dominated by the U and R
fluxes. Such optical continua showing a Balmer jump in emission and stronger
emission in R than in B and V, suggest an optically thin emitter.

\section{The dwarf--nova disc instability model}

\subsection{Quiescent disc in the DIM}

Quiescent dwarf nova discs are
not {\sl stationary} so most formulae familiar from the `standard'
disc model do not apply in this case.

According to the standard version of the DIM a quiescent dwarf nova
disc must be entirely in a cold (low) state. This  means that its
surface density at any point must be lower than the maximum surface
density on the cold `branch' $\Sigma_{\rm max}$.  For $\alpha < 0.5$
this maximum surface density can be fitted by the following formulae
(Hameury et al. 1998)
\begin{equation}
\Sigma_{\rm max}= 13.4\alpha^{-0.83} m_1^{-0.38} r_{10}^{1.14} \ {\rm g
\ cm^{-2}} 
\end{equation} 
The corresponding critical equilibrium accretion rate is
\begin{equation}
\dot M_{\rm crit} = 4\times 10^{15} \alpha^{0.004} m_1^{-0.88} r_{10}^{2.65}
\ {\rm g \ s^{-1}} 
\end{equation}
and the critical effective temperature is given by
\begin{equation}
T_{\rm eff, crit}=5800 \alpha^{0.001} m_1^{0.03} r_{10}^{-0.09} \ {\rm K}
\label{teff}
\end{equation}

where $ m_1$ is the white dwarf mass in solar units, $\alpha$ is the
viscosity parameter and $r_{10}$ is the distance from the center in
$10^{10}$ cm.  The fact that the critical effective temperature depends
very weakly on the parameters defining disc structure, (for a given
convection energy transport prescription) is well known and easy to
explain (see e.g. Cannizzo 1993).

The effective temperature profile ($T_{\rm eff} < T_{\rm eff, crit}$
of the quiescent disc
is, according to the model, rather flat, much flatter than the 
stationary disc profile $ T_{\rm eff} \sim R^{-0.75}$.

Clearly $\dot M \neq const.$ so a DN quiescent disc is not in
equilibrium. This is not surprising, since if it were in a stable
equilibrium it would have no reason to go into outburst. Matter
transferred from the secondary is, therefore, accumulated somewhere in
the disc. The location of the accumulation region depends on the values
of the mass transfer rate and of the viscosity in the disc.

The quiescent disc is not in a {\sl viscous} equilibrium but the {\sl 
thermal} equilibrium condition is very well satisfied (Hameury et al. 1998).
In terms of the S--curve picture it corresponds to
the system point moving, on a (long `secular') viscous time, along the 
lower, cold, branch of the equilibrium curve.

Since  $\dot M \neq const.$ the well known stationary state formulae 
relating viscous heating and effective temperature to the accretion rate
are not longer valid. Their equivalents can be easily derived from the
standard equations describing non--stationary accretion discs..
 
The radial velocity in a Keplerian disc can be written as (see e.g. Frank,
King and Raine 1992)
\begin{equation}
v_{\rm R} = - {3 \over \Sigma R^{1/2}} {\partial \over \partial R}
\left(\nu \Sigma R^{1/2} \right) 
\end{equation}
so that the accretion rate $\dot M \equiv 2 \pi R \Sigma v_{\rm R}$
is
\begin{equation}
\dot M= 3\pi \nu \Sigma \left[ 2 {\partial \ln \left(\nu \Sigma \right) 
\over
\partial \ln R} + 1 \right] 
\end{equation}
The formula for the viscous heating per unit surface $Q^+$ takes then the
form
\begin{equation}
 Q^+ \equiv {9 \over 8} \nu \Sigma \Omega^2=
{3 \over 8 \pi} \left[ 2 {\partial \ln \left(\nu \Sigma \right)
\over
\partial \ln R} + 1 \right]^{-1} {GM \dot M \over R^3}
\end{equation}
Taking into account the inner boundary condition ($\nu \Sigma(R_{\rm in})=0$)
one can write $\nu \Sigma$ as
\begin{equation}
\nu \Sigma \equiv \mu f= \mu\left(t,R \right) 
\left[1 - \left({R_{\rm in}\over R}\right)^{1/2} \right]    
\end{equation}
so that
\begin{equation} 
\dot M= 3\pi \nu \Sigma \left[ 2 {\partial \ln \mu 
\over
\partial \ln R} + f^{-1} \right]  
\end{equation} 
For $\mu = const$ one obtains the familiar formula valid in case
of viscous equilibrium but in the general, non--stationary case the
factor $\left[ 2 {\partial \ln \mu /\partial \ln R} + f^{-1} \right]$
can take values from 2 to 5, say. This is not negligible if one wants
to determine the value of the effective temperature $ T_{\rm eff}=
(Q^+/\sigma_{\rm SB})^{1/4}$ because the ratio of its highest to its
lowest values defining the instability `strip' is smaller than 1.5.

\subsection{The high $\alpha$ problem}

Attempts to `fit' quiescent dwarf nova discs result in very high 
($\alpha \sim 10^2 - 10^3$ values of the viscosity parameters. The 
reason for that can be easily understood.

The energy equation is written in the form:
\begin{equation}
{3\over 2} \Omega \alpha P H = \int F_{\nu}d\nu =\sigma T_{\rm eff}^4  
\label{heating}
\end{equation}
Scaling to characteristic parameters of a quiescent dwarf nova disc, this 
can be written as
\begin{equation}
\alpha \Sigma T  = 6.46 \times 10^6 T_{{\rm eff},5}^4 M_1^{-1/2} 
                   \left({R\over R_{L_1}}\right)^{3/2}
                   \left({R_{L_1}\over R_{\odot}}\right)^{3/2}
\label{highalf}
\end{equation}
where $T_{{\rm eff},5}$ is the effective temperature in units of 5000
K, and the distance has been expressed in units of the distance from
the white dwarf to the $L_1$ point $R_{L_1}$ as used in the MEM
maps.  It is easy to see that for central temperatures $T \lta 10^4
K$, low values of $\Sigma < 1$ required by an optically thin model
imply $\alpha\gg 1$!  This especially the case of isothermal slab
models in which both the lines and the continuum are emitted by the
same medium (Wood et al. 1992).  As we see in Section 5, even a more
refined model with temperature stratification still requires, in the
case of HT Cas, $\alpha\gta 1$.

\section{Models and results}      

\subsection{The `$S$-curves'}

The SW model of  vertical accretion
disc structure assumes that the disc is in a steady state.
Since quiescent dwarf nova discs are in thermal equilibrium one can use
the SW model at each ring of the disc, using the local value of the accretion
rate and taking into account non--steady disc relations described in 
Section 2.2.

\begin{figure}
\begin{center}
\epsfig{figure=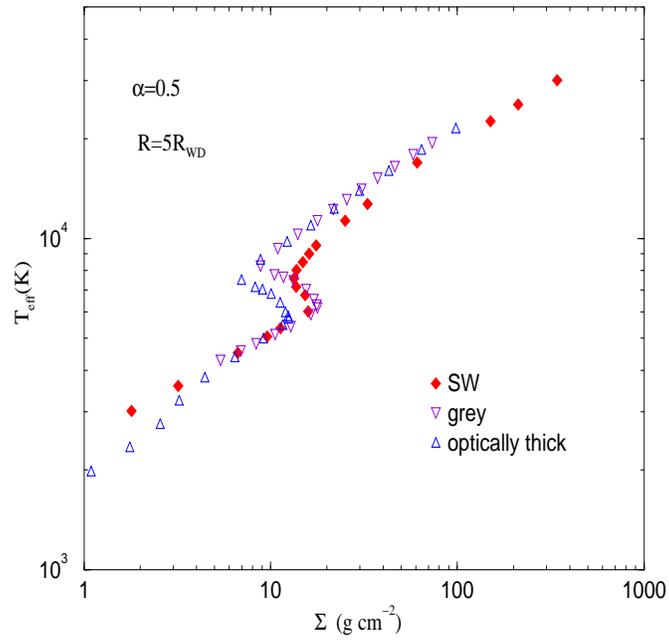,width=10cm, height=10cm}
\caption{The the equilibrium solution $S$-curves for an accretion disc with 
$M_{\rm WD}=0.6 M_{\odot}$, $\alpha=0.5$ at $R=5 R_{\rm WD}$. Diamonds
correspond to models calculated with the Shaviv \& Wehrse model, down
pointing triangles show disc solutions found with the Hameury et al. model
in the `grey' approximation and  triangles up correspond to
diffusion (`optically thick') approximation.}
\label{scurve}   
\end{center}
\end{figure}

First, we verified that SW thermal equilibrium accretion disc solutions
have the same general properties as the disc models used in the DIM.
Using the SW model we calculated series of models for several values of
the radius and the viscosity parameter $\alpha$. We plotted this
solutions on a customary $\Sigma - T_{\rm eff}$ diagramme. An example
is shown on Figure (\ref{scurve}). We compare the SW models with the
ones obtained with the Hameury et al. (1998) code. In this code two
approximations can be used: either the transfer is calculated in the
diffusion approximation or a (simplified) `grey atmosphere' approach is
adopted. Fig. (\ref{scurve}) show that there is very good agreement
between SW and `grey' models on the cold branch of solutions ($\Sigma
\leq \Sigma_{\rm max}$) but there are deviations on the unstable and
hot branches. The SW and Hameury et al. (1998) approach give the same
results only for $T_{\rm eff} \gta 20 000$ K.  This differences are
most probably due mainly to different opacity tables used in the two
codes (Idan et al. 1998).  Since in the standard DIM one has to assume a
`jump' in $\alpha$ (see e.g. Hameury et al. 1998) the small differences in
the upper branches of solutions are of no practical importance for the
description of a DN outburst. The excellent agreement on the lower
branch makes it consistent to model emission from quiescent dwarf nova discs
by using in the SW code $T_{\rm eff}(r)$ obtained from the DIM.

\subsection{The quiescent disc in HT Cas}

We used the SW code to study emission properties of the quiescent 
accretion disc in HT Cas.
We did not try to fit the brightness temperature distribution obtained
by Wood et al. (1992). First, it is not clear what is the sense of a procedure
in which one first fits intensity maps to eclipse curves by using MEM,
then, in a second step, one fits the physical parameters to the intensities
(Horne 1993; Vrielmann 1997). Second, the status of the so-called `Physical 
Parameter Eclipse Mapping' proposed by Vrielmann (1997) is not well
established and its application to HT Cas produced results which are
in contradiction both with Wood et al. (1992) and the fundamental assumption
of the DIM. In any case using the SW code for the `Physical Parameter
Eclipse Mapping' would require a prohibitive amount of computer time.

We adopted, therefore, a simple and pragmatic approach and just tried
to find the range of accretion rates and values of the viscosity parameter 
which gives the best description of the observed properties of the quiescent 
disc in HT Cas. We used as a guide the total disc colours which are not 
determined by the MEM fitting procedure (but they are model dependent).
We tried to reproduce flux ratios of the U,B,V and R
wavelengths (folded through bandpasses response functions of the
Stienning photometer). This procedure is independent of the distance
to HT Cas. In this way we find a range of $\dot M$ and $\alpha$ values which 
correspond to flux ratios of both the total disc emission and
the reconstructed radial distribution.

Our main aim was to check if observed properties of the quiescent
disc in HT Cas are consistent with the general assumptions of accretion disc
models. Results of previous studies (Wood et al. 1992; Vrielmann
1997) were in a flagrant contradiction with the standard disc model (not
only with the DIM, so that constraints obtained on critical temperatures
etc. do not have much meaning). Here we tried to see if an improved
model of the disc vertical structure may help to reconcile models
and observations.

The total disc fluxes in UBVR (which are just slightly different form
Johnson's standards) in HT Cas are (Wood et al. 1992):  $F_U=0.71\pm
0.02$, $F_B=0.51\pm 0.03$, $F_V=0.5\pm 0.05$, $F_R=0.8 \pm 0.05$ (in
millijanskys). Clearly the Balmer jump is in emission and $F_R > F_B,
F_V$ confirms that the continuum is emitted by an optically thin
medium.  Emission lines apparently contribute $\sim 30\%$ to the $R$
emission, but even with this contribution subtracted, the $R$ emission
is still high compared with fluxes in shorter wavelengths. The fluxes
given above assume $E(B-V)=0$ but since the reddening to HT Cas is not
well determined, we have also used $E(B-V)=0.2$ which, probably, is the
maximum value allowed by observations (Wood et al. 1992).

First, we studied the  disc flux ratios at given radius as a function
of the accretion rate for a given $\alpha$. An example is shown on
Figs. (2) and (3) for $R=5R_{\rm WD}$.  We assumed that, 
the disc effective temperature is constant, parallel to the critical 
effective temperature (we can neglect in Eq. 3 the weak
dependence on $R$ and $\alpha$), so that in the disc $\dot M \sim R^3$.
(In what follows we use $\dot M = \dot M_0 \left(R/9R_{\rm WD}\right)$).
This is a good approximation of the temperature profiles obtained in
the framework of the DIM.
One can see that the U fluxes ratios all have a minimum at accretion
rates $\dot M_0 \sim 10^{15}-10^{16}  $ g s$^{-1}$.  
($\dot M_0$ is the accretion rate
corresponding to a steady disc, the real accretion rate in this case is
7 times larger - see 3.1. Eq. 5). This feature is almost independent of the
value of $\alpha$. At approximatly the same value of accretion rate 
the $F_R/F_V$, $F_V/F_B$ and $F_R/F_B$ ratios intersect. Clearly, 
this range of
values of the accretion rates corresponds to the transition between an
optically thick (for higher $\dot M$'s) and optically thin regime (for
lower  $\dot M$'s). One can see that observed flux ratio correspond
clearly to an optically thin disc. We found that $\dot M_0 \approx 
1.6 \times 10^{15}$ and $\alpha \approx 1.5$ are the best choice
of parameters.

We use then the total disc colours to check if this choice is consistent
with this set of observations.

Figures (4),(5),(6) and (7) show the calculated flux
ratios as a function of $\alpha$. The $F_U/F_B$, $F_U/F_V$ and
$F_U/F_R$ ratios depend rather strongly on $\alpha$ whereas $F_V/F_B$,
$F_R/F_B$ and $F_R/F_V$ show only a weak dependence on this parameter.
The maximum in the $F_U$ ratios is due to the fact that with increasing
$\alpha$ the midplane temperature also increases but the optical depth
decreases, so that for some value of the viscosity parameter the latter
effect begins to dominate. We increased $F_R$ by 30\% in order to
account for the line emission that has not been included in the
vertical structure model.

One can see that if $E(B-V)=0$, calculated $F_V/F_B$ and $F_R/F_V$ are
outside the observed $\sim 90\%$ confidence (2$\sigma$) limit. For  
$E(B-V)=0.2$ one gets a very good agreement between the model and
observations especially for the $F_V/F_B$, and $F_R/F_V$ and $F_R/F_B$
ratios, but they depend only weakly on $\alpha$. 

We conclude, that if the quiescent disc in HT Cas is unsteady and its
effective temperature profile is similar to the one predicted by
the DIM, the required value of $\alpha$ is marginally consistent
with the assumptions of the standard accretion disc model but incompatible
with values required to obtain dwarf nova outbursts in this model framework.
The value of the mass transfer rate $\gta 10^{-10}M_{\odot}$ y$^{-1}$
is consistent with the, very uncertain, distance determinations (Wood
et al. 1992).

\section{Conclusions}

We showed that the properties of equilibrium accretion disc configurations
which are the basis of dwarf nova instability model (`$S-curves$') are not 
drastically  modified by using the Shaviv \& Wehrse radiation transport code
to calculate disc equilibria.

The use of this code to describe emission from the quiescent disc of
dwarf nova HT Cas leads to values of the viscosity parameter $\alpha$
which are larger than unity. They are lower than the large values
obtained in some previous studies but still too high to be in agreement
with the fundamental assumptions of the standard disc model and it is
doubtful that additional refinements would change $\alpha$ by more than
a factor 2. In any case the main problem, is that quiescent dwarf nova
discs seem to be optically thin while the model seem to require
optically thick media.  The reason is that, according to the model, the
ratio between the quiescent and outburst values of the viscosity
parameter $\alpha$ must be $\sim 4\ - \ 10$ and the outburst values are
$\alpha_h \sim 0.1$.  The quiescent value should be  then $\alpha_c
\sim 0.01$, a value which is also suggested by dwarf nova recurrence
times (Livio \& Spruit 1991). The DIM code with $\alpha_c \sim 0.5$
and $\alpha_h \sim 2.5$ gives light curves which are rather dissimilar
to those of HT Cas. One should remember, however, that such light
curves cannot be reproduced by the standard DIM anyway (Lasota et al.
1995).  The possibility of truncated disc in HT Cas should be studied
in more details. At the first sight, the reconstructed light-curves do
not seem to reflect such a feature (Smak 1994) but the significance of
reconstructed points near the white dwarf is not clear (see e.g
Vrielmann 1997). Observations which suggest a rather small X-ray
emitting region (Mukai et al. 1997) are {\sl not} in contradiction with
the existence of a truncated disc.

It has been speculated that the optically thin emission form quiescent
dwarf nova disc could have its origin in a `chromosphere' or a
`corona'.  A corona (Shaviv \& Wehrse 1986), i.e. a medium at $T\gta
10^6$ K cannot produce the observed colours. A chromosphere could be
responsible for line emission but it is difficult to see how such a
narrow structure could emit a continuum with the observed properties.

Finally, one cannot exclude that the picture of a `smooth' homogenous
disc is not an adequate description of  quiescent dwarf nova accretion
flows.

If HT Cas is indeed the `Rosetta stone' of dwarf nova models we have
rather small chances to decipher its hieroglyphs.

\section*{Acknowledgments} I.I. and JPL thank Joe Smak for very helpful
discussions. We thank Janet Wood for useful informations about observations
of HT Cas.

\section*{References}

\noindent Cannizzo, J.K. 1993, in {\it Accretion Disks in Compact Stellar 
           Systems}, ed. J.C. Wheeler, (Singapore: World Scientific), p. 6

\noindent Cannizzo, J.K. 1997, in {Cataclysmic Variables and Related Objects},
     Proceedings of the 13th North-American Workshop; ed. E. Kuulkers et al. ,
     ASP Conference Series, in press

\noindent Cannizzo, J.K. \& King, A.R. 1998, ApJ, in press

\noindent Dumont, A.-M. Collin-Souffrin, S., King, A.R. \& Lasota, J.-P. 1991,
                         A\&A, 242, 503

\noindent Frank, J., King, A.R. \& Raine, D. 1992, {\it Accretion Power
in Astrophysics}, (Cambridge: CUP)

\noindent Gammie, C. \& Menou, K. 1997, ApJ, 492, L75

\noindent Hameury, J.-M.,  Lasota, J.-P. \& Hur\'e, J.-M., 1997, MNRAS
 
\noindent Hameury, J.-M., Menou, K., Dubus, G., Lasota, J.-P. \& 
                                    Hur\'e, J.-M. 1998, MNRAS,  in press

\noindent Horne, K. 1993, in {\it Accretion Disks in Compact Stellar 
          Systems}, ed. J.C. Wheeler, (Singapore: World Scientific), p. 117

\noindent Horne, K., Wood, J.H., Stienning, R.F. 1991, ApJ 378, 271

\noindent Idan, I. \& Shaviv, G. 1996, MNRAS 281, 604

\noindent Idan, I. et al. 1998, in preparation

\noindent King, A.R. 1997, in {\it Relativistic Gravitation and Gravitational
       Radiation} ed. J.-A. Marck \& J.-P. Lasota, (Cambridge: CUP),
                    p. 105

\noindent Lasota, J.-P. 1996, in {\it Cataclysmic Variables and Related 
    Objects}, IAU Coll. 158, ed. J.H. Wood et al. (Dordrecht: Kluwer), p. 385

\noindent Lasota, J.-P., Hameury, J.-M. \& Hur\'e, J.-M., 1995, A\&A, 302, L29

\noindent Lin, D.N.C., Williams, R.E. \& Stover, R.J. 1988, ApJ, 327, 234

\noindent Livio, M. \& Spruit, H. 1991, A\&A, 252, 189

\noindent Livio, M. \& Pringle J., 1992, MNRAS 259, 23p


\noindent Marsh, T. R. 1987, MNRAS, 228, 779

\noindent Meyer F. \& Meyer-Hofmeister E. 1994, A\&A 288, 175

\noindent Mineshige, S. \& Wood, J.H. 1989, MNRAS, 241, 259

\noindent Mukai, K., Wood, J.H., Naylor, T., Schlegel, E.M. \& Swank, J.H. 1997, 
ApJ, 475, 812

\noindent Patterson, J. 1981, ApJS, 45, 517

\noindent Rutten, R.G.M.R., van Paradijs,J. \& Tinbergen, J. 1992, A\&A, 
260, 213

\noindent Shaviv, G. \& Wehrse, R. 1986, A\&A, 169, L5

\noindent Shaviv, G. \& Wehrse, R. 1991, A\&A, 251, 117

\noindent Smak J. 1984, Acta astron., 34, 161

\noindent Smak J. 1994, Acta astron., 44, 265

\noindent Smak J. 1996, in {\it Cataclysmic Variables and Related Objects},
IAU Coll. 158,  ed. J.H. Wood et al. (Dordrecht: Kluwer),  p. 45

\noindent Smak J. 1998, this volume

\noindent Tylenda, R. 1981, Acta astron., 31, 127

\noindent Vrielmann, S. 1997, PhD Thesis, University of G\"ottingen 

\noindent Wade, R.A. 1988, ApJ, 335, 394

\noindent Warner, B. 1995, {\it Cataclysmic Variable Stars}, (Cambridge: CUP)

\noindent Wenzel, W. 1987, Astron. Nachr., 308, 75

\noindent Williams, G.A. 1991, AJ, 101, 1929

\noindent Williams, R.E. 1980, ApJ, 235, 939

\noindent Wood, J.H. 1990, MNRAS, 243, 219

\noindent Wood, J.H., Horne, K. \& Vennes, S. 1992, ApJ 385, 294

\noindent Wood, J.H., Horne, K., Berriman, G. \& Wade, R. 1989, ApJ, 341, 974

\noindent Wood, J.H., Naylor, T., Hassal, B.J.M. \& Ramsayer, T.F. 1995, 
MNRAS, 273, 772

\noindent Wood, J.H., Horne, K., Berriman, G., Wade, R. O'Donoghue, D.
\& Warner, B. 1986, MNRAS, 219, 629

\noindent Zhang, E.-H., Robinson, E.L. \& Nather, R.E. 1986, ApJ, 305, 740

\vfill\eject

\vfill\eject
\begin{figure}
\begin{center}
\epsfig{figure= 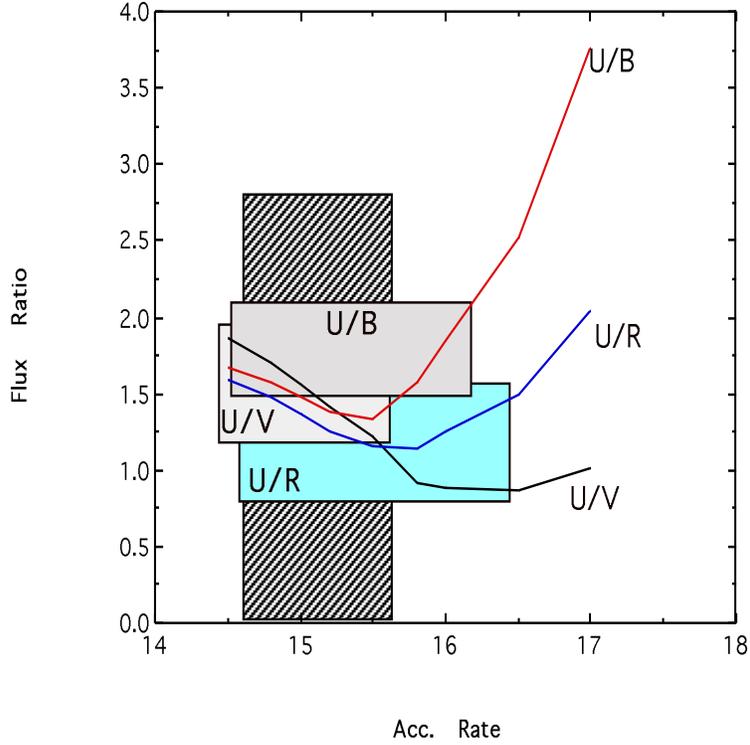,width=10cm, height=10cm}
\caption{$F_U/F_B$, $F_U/F_V$ and $F_U/F_R$ flux ratios as
functions of the accretion rate for $R=5R_{\rm WD}$, and $E(B-V)$=0. The
grey zones correspond to the 'observed values' ($\pm 1 \sigma$) taken
from the MEM reconstruction.  The lower $\dot M$ limit on calculated
flux ratios results from the lower limit on the temperature in the
opacity tables used. The grey `boxes' extend to the highest value of
the accretion compatible with `observations'.  Boxes are slightly
shifted for clarity.  The dashed area corresponds to the range of
accretion rates where best agreement between model and `observations'
can be found.}
\label{radcol1} 
\end{center}
\end{figure}

\vfill\eject
\begin{figure}
\begin{center}
\epsfig{figure= 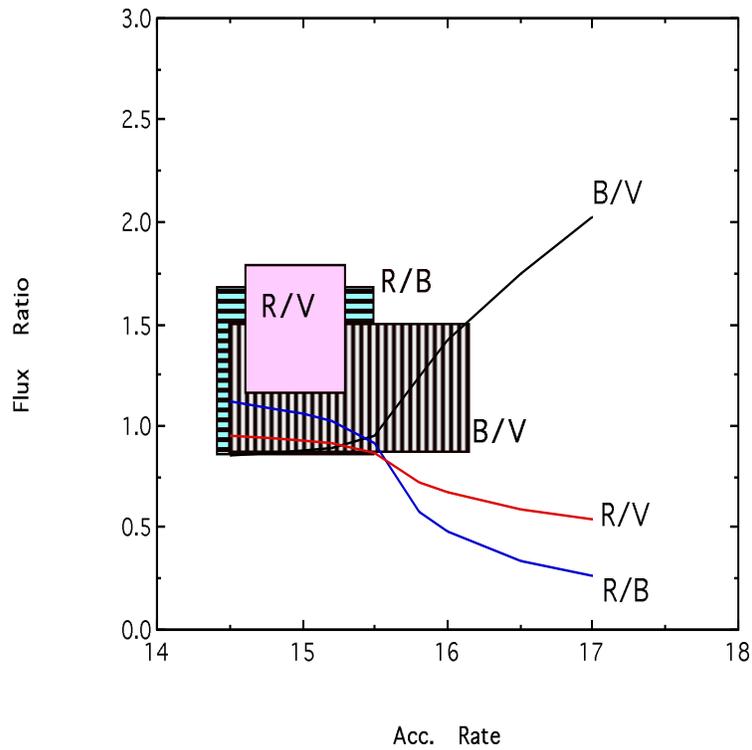,width=10cm, height=10cm}
\caption{The same as in Fig. 2 but for $F_R/F_V$, $F_R/F_B$ and $F_V/F_B$.
Here the calculated $F_R/F_V$ ratio is always out of the observed range, but
taking a $2 \sigma$ limit and/or adding some reddening would easily solve the
problem and give values of  acceptable $\dot M$'s close to the one in Fig. 2}
\label{radcol2} 
\end{center}
\end{figure}

\vfill\eject
\begin{figure}
\begin{center}
\epsfig{figure= 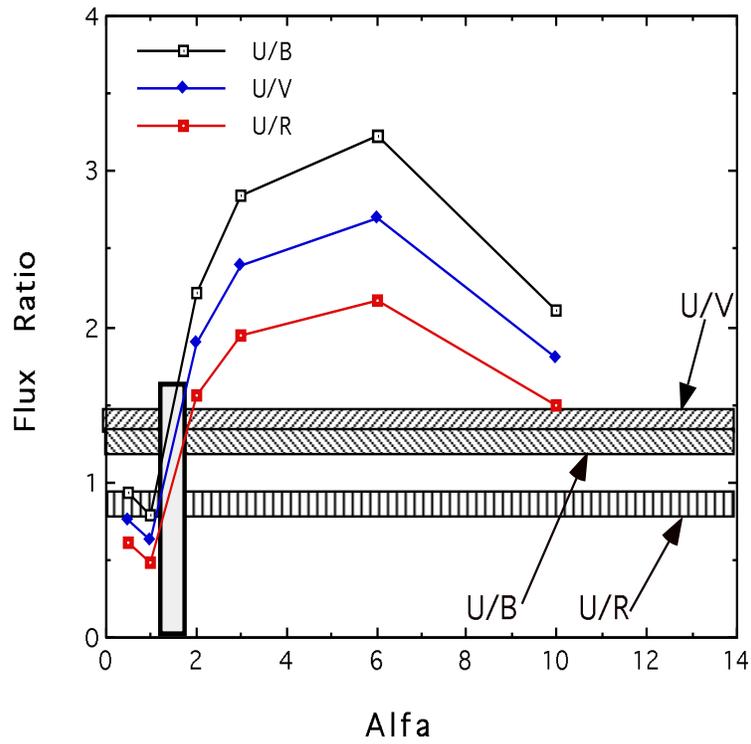,width=10cm, height=10cm}
\caption{Total disc flux ratios $F_U/F_B$, $F_U/F_V$ and $F_U/F_R$ as  
functions of $\alpha$. The observed flux ratios are represented by 
$1 \sigma$ dashed bands. The grey rectangle corresponds to values of
$\alpha$ for which best agreement between model and observations 
can be found.
$E(B-V)=0$}
\label{cratio1}
\end{center}
\end{figure}

\vfill\eject
\begin{figure}
\begin{center}
\epsfig{figure= 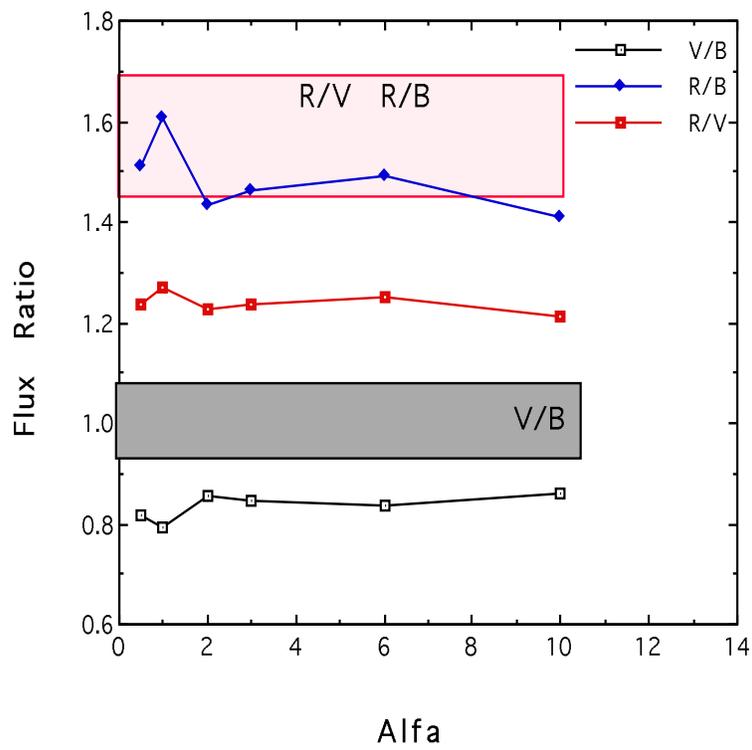,width=10cm, height=10cm}
\caption{The same as in Fig. 4 but for $F_R/F_V$, $F_R/F_B$ and $F_V/F_B$.
}
\label{cratio2}
\end{center}
\end{figure}

\vfill\eject
\begin{figure}
\begin{center}
\epsfig{figure= 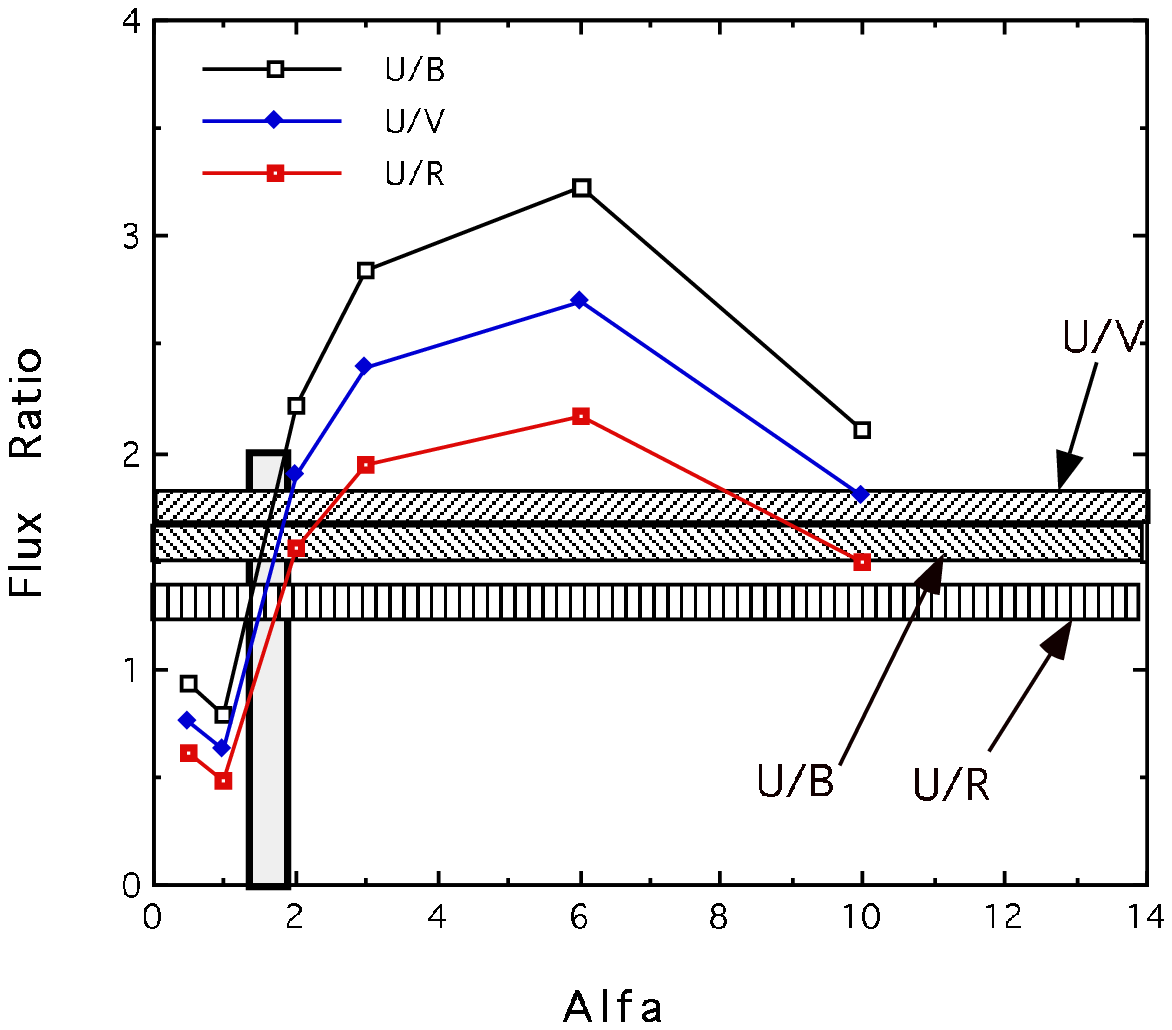,width=10cm, height=10cm}
\caption{The same as Fig. 4 but with $E(B-V)=0.2$}
\label{cratio1.022} 
\end{center}
\end{figure}

\vfill\eject
\begin{figure}
\begin{center}
\epsfig{figure= 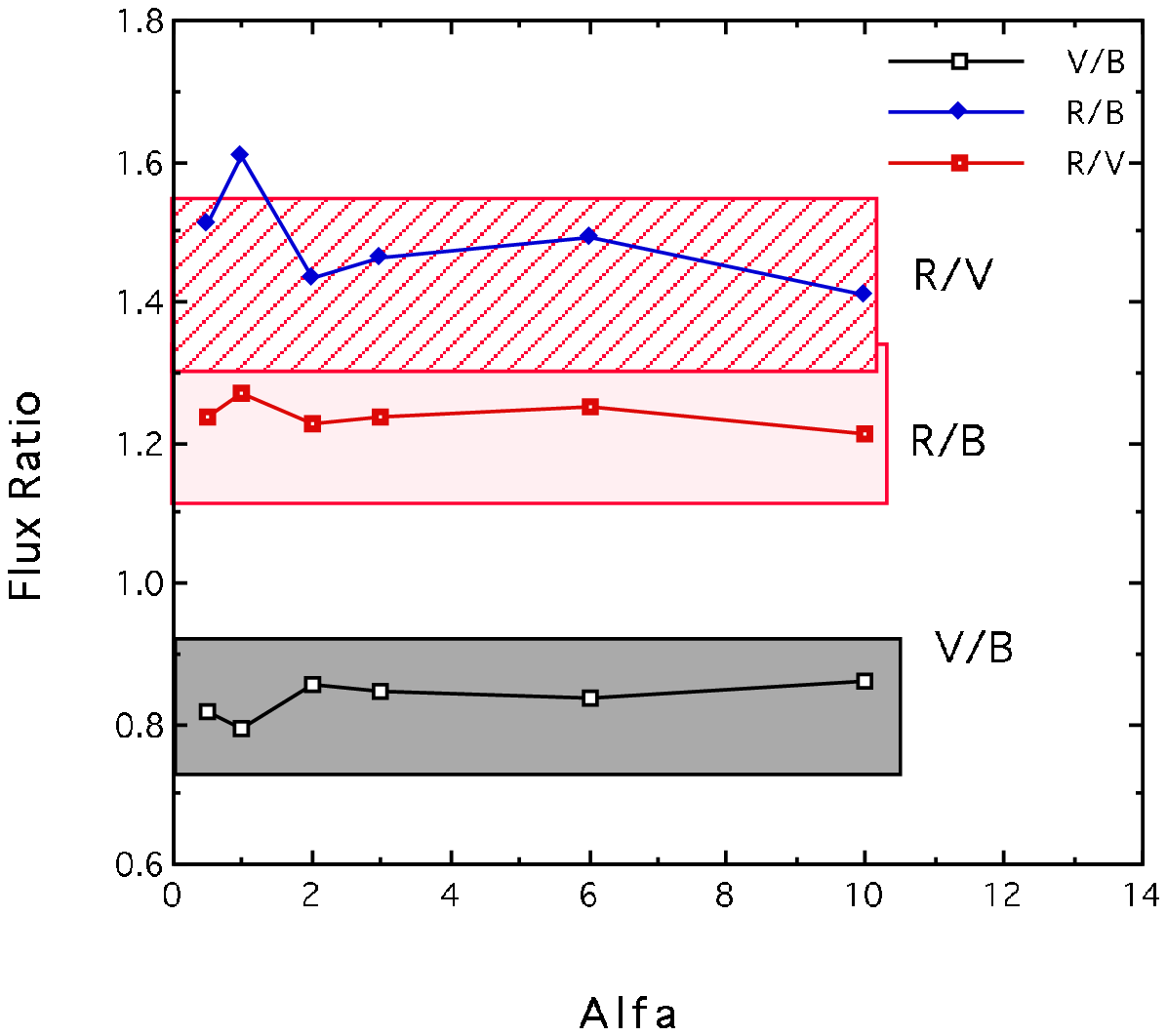,width=10cm, height=10cm}
\caption{The same as Fig. 5 but with $E(B-V)=0.2$}
\label{cratio2.022} 
\end{center}
\end{figure}

\end{document}